\documentclass[twocolumn]{aastex631}

\shorttitle{}
\shortauthors{}
\graphicspath{{./}{figures/}}

\begin{document}

\title{A New Polar Ring Galaxy Discovered in the COSMOS Field}

\author[0000-0003-1966-5742]{Minoru Nishimura}
\affiliation{The Open University of Japan \\
2-11 Wakaba, Mihama-ku \\
Chiba, Chiba, Japan}

\author[0000-0001-6473-5100]{Kazuya Matsubayashi}
\affiliation{Institute of Astronomy, The University of Tokyo \\
2-21-1 Osawa \\
Mitaka, Tokyo, Japan}

\author[0000-0001-5211-7807]{Takashi Murayama}
\affiliation{Astronomical Institute, Graduate School of Science, Tohoku University\\
Aramaki, Aoba\\
Sendai, Miyagi, Japan}

\author[0000-0003-2247-3741]{Yoshiaki Taniguchi}
\affiliation{The Open University of Japan \\
2-11 Wakaba, Mihama-ku \\
Chiba, Chiba, Japan}



\begin{abstract}

  In order to understand the formation and evolution of galaxies fully,
  it is important to study their three-dimensional gravitational potential for a large sample of galaxies.
 Since polar-ring galaxies (PRGs)  provide useful laboratories for this investigation, we have started our detailed study 
 of a sample of known PRGs by using the data set obtained by the Hyper Suprime-Cam Subaru Strategic Program (HSC-SSP).
 During the course of this study, we have discovered a new PRG,
  identified as SDSS J095351.58+012036.1. Its photometric redshift is estimated as $z \sim 0.2$. 
  The polar ring structure in this PRG appears to be almost perpendicular to the disk of its host galaxy
  without any disturbed features. Therefore, this PRG will provide us with useful information
  on the formation of such an undisturbed polar structure.
  We discuss its photometric properties in detail.

\end{abstract}



\section{Introduction}

Polar ring galaxies (PRGs) are galaxies with a polar structure (either a ring or a disk of gas, dust, and stars) 
rotating in a plane almost perpendicular to the major axis of its host galaxy.
The archetypical  PRG is NGC 2685 (or Arp 336).
\citet{sandage_hubble_1961} gives the following note on this galaxy;
^^ ^^ There are two axes of symmetry for the projected image; most galaxies have only one." 
Its host galaxy is an  S0 galaxy, with the polar structure rotating through the minor axis of
 the host galaxy \citep{schechter_ngc_1978}. From a theoretical point of view, 
it is considered that the stable maintenance of the polar structure is due to the precessional motion of the polar structure \citep{steiman-cameron_stable_1982}.

So far, more than 400 candidates of  PRGs have been discovered  to date (e.g., \cite{whitmore_cnew_1990}, \cite{moiseev_cnew_2011}).
However,  among them, only dozens have been confirmed as real PRGs by spectroscopic observations (e.g., \cite{egorov_metallicity_2019}). 

Since most galaxies reside in large-scale structures, interactions or mergers among galaxies are not rare events for them.
In order to understand the dynamical and morphological evolution of galaxies, it is important to study how such interactions or mergers
affect the evolution of galaxies by using  large unbiased samples of PRGs.
Indeed,  observational properties of  PRGs  allow us to investigate a wide range of issues related to their galaxy formation and evolution:
 e.g., the baryonic matter accretion (e.g., \cite{egorov_metallicity_2019, smirnov_active_2020}), the rate of galaxy interactions or mergers (e.g., \cite{reshetnikov_polar-ring_2011, reshetnikov_galaxies_2019}), and the 3D distribution of mass in the dark halo (e.g., \cite{khoperskov_be_2014, zasov_dark_2017}).
In typical PRGs, their host galaxies are mostly early-type galaxies (E/S0), while their polar structures are generally young, blue, and gas rich \citep{reshetnikov_new_2019}. 
In addition, the observed polar structures show a wide variety in their morphology; e.g., a narrow ring or a wide annulus (\cite{whitmore_few_1991}), 
a spindle-, a Saturn-,  or a worm-like structure \citep{faundez-abans_morphology_1998} , inner polar structure \citep{moiseev_inner_2012}.

As for the formation of PRGs, the following three mechanisms have been theoretically proposed;
 galaxy mergers (e.g., \cite{bekki_formation_1997, bekki_formation_1998, bournaud_formation_2003}), 
 accretion of matter from an approaching galaxy (e.g., \cite{reshetnikov_global_1997, bournaud_formation_2003}), 
 and cold accretion from filaments in intergalactic space (e.g., \cite{maccio_origin_2006, brook_formation_2008, snaith_halo_2012}).
However, it is still uncertain which mechanism is important for the formation of PRGs.

In order to improve the understanding of the formation and evolution of PRGs,
we have started our systematic search for PRG candidates using the data set 
obtained by the Hyper Suprime-Cam Subaru Strategic Program (HSC-SSP; \cite{aihara_hyper_2018}).
During the course of this search, 
we have discovered a new PRG candidate SDSS J095351.58+012036.1  (hereafter J0953) .
This galaxy is located 
 at the edge of the Cosmic Evolution Survey (COSMOS; \cite{scoville_cosmic_2007}) field. 
 This survey covers a 2 square degree field. It is designed to probe the galaxy formation and evolution 
 as functions of both cosmic time (redshift) and the local galaxy environment. 
It is noted that only one PRG candidate has been 
  found in the COSMOS Field; SPRC 093   \citep{moiseev_cnew_2011}.

In this paper, we discuss the observational properties of our new PRG candidate J0953. 
Throughout this paper, we use the following cosmological parameters; 
$H_0$ = 70 km s$^{-1}$ Mpc$^{-1}$, $\Omega_{\rm m}$ = 0.3, and $\Omega_{\Lambda}$ = 0.7.
All magnitudes given in this paper are in the AB magnitude system.

\section{SDSS J095351.58+012036.1}

   The new PRG candidate in our study, J0953, is found at the  eastern edge of the COSMOS field in Wide-layer images from
   the second Public Data Release (PDR2; \cite{aihara_second_2019}) of HSC-SSP. 
The HSC-SSP survey consists of the following three layers: Wide, Deep, and UltraDeep.
The Wide layer covers about 300 deg$^2$ in all the five broad-band filters ($grizy$)
to the nominal survey exposure (10~min in $gr$ and 20~min in $izy$). 

The HSC-SSP PDR2 data have already been processed with the data reduction pipeline hscPipe version 6 (\cite{bosch_hyper_2018},
\cite{aihara_second_2019}),  including the following procedures; the point spread function (PSF) modeling, object detection, and photometry. 

\begin{figure}
   \begin{center}
      \includegraphics[width=80mm]{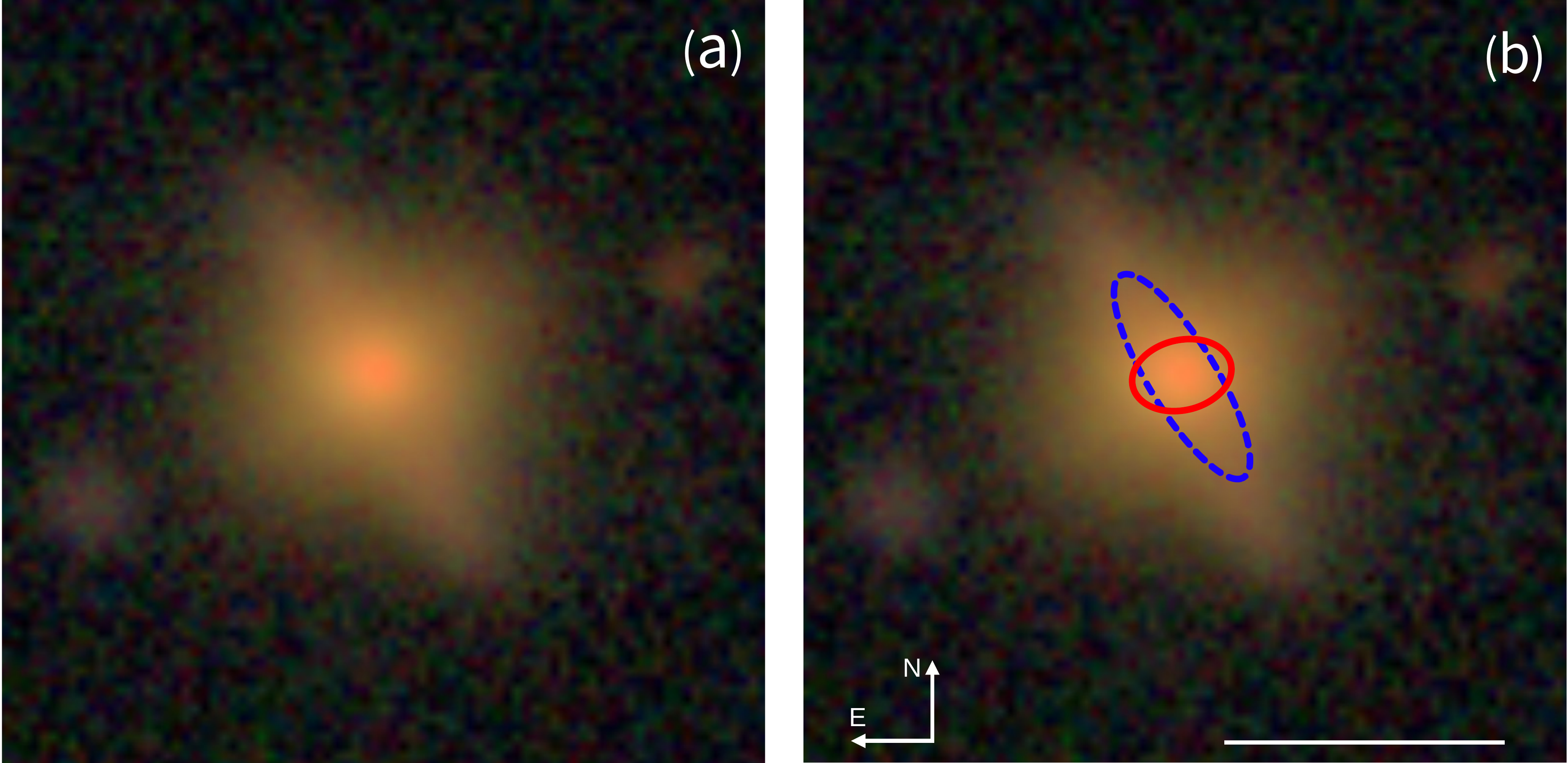}
   \end{center}
   \caption{The $gri$ composite image of J0953 taken from the hscMap
    (Dataset: PDR2 WIDE, Mixer: SDSS TRUE COLOR). North is up and east is left. The white bar displayed in panel (b) corresponds to the angular size of 5 arcsec, or $\approx$ 24 kpc at $z$ = 0.2 (see text). The solid and dashed lines in panel (b) show the half-light radii of the host galaxy and polar structure, respectively.
   }
   \label{fig:5image}
\end{figure}

In figure \ref{fig:5image} (a), we show the $g$-, $r$-, and $i$-band composite image of J0953
taken by hscMap tool\footnote{hscMap: $\langle $https://hsc-release.mtk.nao.ac.jp/hscMap-pdr2/app/$\rangle $} (\cite{koike_hscmap_2019}) with SDSS TRUE COLOR mixer. 
Apparently, a polar structure or ring can be seen in an NE-SW direction
almost perpendicular to the disk of the host galaxy.

J0953 is identified as a galaxy by the Sloan Digital Sky Survey (SDSS; \cite{york_sloan_2000}); SDSS J095351.58+012036.1.
Its photometric properties are reported as follows;
$u^{\prime}=21.49\pm 0.25$ mag, $g^{\prime}=19.73\pm 0.03$ mag, 
$r^{\prime}=18.83\pm 0.02$ mag, $i^{\prime}=18.34\pm 0.025$ mag, and $z^{\prime}=18.02\pm 0.05$ mag.
These data give a  photometric redshift of $0.146\pm 0.036$ \citep{beck_photometric_2016}.
Note that there is no spectroscopic observation of J0953 and thus no spectroscopic redshift is available.

In table \ref{tab:data} we present the photometric properties of J0953 taken from the HSC-SSP database by 
CAS search tool\footnote{CAS search: $\langle $https://hsc-release.mtk.nao.ac.jp/datasearch/$\rangle $}.
It is noted that the magnitudes given in this table are slightly brighter than the SDSS magnitudes given above.
We consider that this slight difference is probably due to the deeper imaging of the HSC-SSP Wide layer.
Using these photometric data, the HSC-SSP group estimates photometric redshifts of J0953 using the two codes.
 One is
 the DEmP code (\cite{hsieh_estimating_2014}), and the other is the MIZUKI code (\cite{tanaka_photometric_2015}). 
 These two codes give $z = 0.19$ (DEmP) and  $z = 0.20$ (MIZUKI), respectively.
 Since these are consistent within their errors,
we adopt the photometric redshift of J0953, $z = 0.20$ in this paper.
This gives the luminosity distance of J0953, 980 Mpc. 

\begin{deluxetable*}{cc}
  \tablenum{1}
  \tablecaption{Properties of J0953 from the HSC-SSP Database\label{tab:data}}
  \tablewidth{0pt}
  \tablehead{
  \colhead{Property Name} & \colhead{Data}
  }
  \startdata
  RA (J2000.0)  & 09h53m51s.59 \\
  Dec (J2000.0) & +01D20'36".3 \\
  $g$ & $19.43$ \\
  $r$ & $18.58$ \\
  $i$ & $18.12$ \\
  $z$ & $17.75$ \\
  $y$ & $17.65$ \\
  Photometric redshift (MIZUKI) & $0.20^{+0.09}_{-0.06}$   \\
  Photometric redshift (DEmP)   & $0.19^{+0.05}_{-0.05}$   \\
  Stellar mass (MIZUKI)   & $3.85^{+3.10}_{-2.06} \times 10^{10}$ $M_\odot$  \\
	Star formation rate (MIZUKI)   & $2.66^{+7.77}_{-2.32}$ $M_{\odot}$ y$^{-1}$   \\
  Interstellar absorption: $A_V$ (MIZUKI) & $0.65^{+0.22}_{-0.33}$  \\
  \enddata
  \tablecomments{We note that the photometric errors 
  in all the five bands are less than 0.01 mag.}
  \end{deluxetable*}

We note that
there are neither  companion galaxies nor satellite galaxies around J0953.
In figure \ref{fig:5image}, however, there appear 
two faint objects about 5.7 arcsec in both east and west of J0953.
Based on their photometric properties, they  are background galaxies
with  photometric redshifts of $z \sim 1.3$, respectively.
 
\section{Results and Discussion}

We discuss the possibility that J0953 is the PRG. 
To do so, we need to examine whether at least one of the two components that appear to be vertically intersecting the host galaxy or polar structure is the disk or the ring.

First, PSFs of the $g, r, i, z,$ and $y$ bands  of J0953 are obtained by using the PSF picker\footnote{PSF picker: $\langle $https://hsc-release.mtk.nao.ac.jp/psf/pdr2/$\rangle $}. Next, the FWHM values of these PSFs are  obtained by using the IRAF\footnote{IRAF is distributed by the National Optical Astronomy Observatories, which are operated by the Association of Universities for Research in Astronomy, Inc., under cooperative agreement with the National Science Foundation.} task IMEXAMINE.

The PSF of each band in arcsec is as follows;
0.75 ($g$),
0.85 ($r$),
0.63 ($i$),
0.78 ($z$), and
0.64 ($y$).

Since the PSF size is the largest in the $r$ band,
the other band images are convolved with Gaussian kernels to fit to the $r$-band one
by using the IRAF task GAUSS.
In the subsequent analyses, we use these smoothed images.

In order to investigate  structural properties of both the polar ring and its host galaxy,
we have made the  $g$-, $r$-, $i$-combined image of J0953 to improve the signal-to-noise ratio.
Then, we have carried out the two-component fitting for both the polar ring and the host galaxy,
each of which has a S\'{e}rsic component. In this analysis, we use
the GaLight package\footnote{GaLight is a Python-based open-source package for 2D model fitting of galaxy images in cooperation with lenstronomy (\cite{birrer_lenstronomy_2018}; \cite{birrer_lenstronomy_2021}.) } (version 0.1.6:  \cite{ding_mass_2020}; \cite{ding_galaxy_2021}).

The fitting results are summarized in table \ref{tab:parameters}.
The obtained ellipses with the half-light radius are shown in figure \ref{fig:5image} (b).

The S\'{e}rsic index for the host galaxy is 2.94, suggesting that the host galaxy has an elliptical galaxy-like structure rather than an exponential disk. 
However, \citet{vika_megamorph_2015} reports that the average of the S\'{e}rsic index for the Sab-Sb galaxies is 2.9. Therefore, it is also possible that the host galaxy is a disk galaxy.

On the other hand,
the polar structure has the flatter light profile than the exponential disk; 
its S\'{e}rsic index is 0.47, being much less than 1.
This small S\'{e}rsic index is consistent with that the polar component has a ring-like structure.
Further imaging observation with a higher spatial resolution is necessary to confirm if the polar structure is a ring or not.

\begin{deluxetable*}{lcc}
  \tablenum{2}
  \tablecaption{Structural parameters \tablenotemark{a} \label{tab:parameters}}
  \tablewidth{0pt}
  \tablehead{
   \colhead{Parameter Name} & \colhead{Host galaxy} & \colhead{Polar structure}
  }
  \startdata
     effective radius [$\arcsec$] & $0.89$ & $2.12^{+0.01}_{-0.01}$ \\
     S\'{e}rsic index & $2.94^{+0.02}_{-0.02}$ & $0.47$ \\
     position angle [$\arcdeg$] & $102.2^{+0.5}_{-0.5}$ & $31.4^{+0.1}_{-0.1}$ \\
     axis ratio\tablenotemark{b} & $0.71$ & $0.28$ \\ 
  \enddata
  \tablenotetext{a}{The structural parameters were estimated on the combined image of the $g$, $r$, and $i$ images by GaLight.}
  \tablenotetext{b}{The axis ratio is the semi-minor  to semi-major axis ratio.}
  \tablecomments{No error means that the error is less than 0.01 in each case.}
\end{deluxetable*}

Next, with the structure parameters fixed as those in table \ref{tab:parameters},
we have performed model decomposition by using GaLight in each of the $g$, $r$, $i$, $z$, and $y$ images of J0953
to measure apparent magnitudes of each component.
Table \ref{tab:magnitude} lists the  apparent magnitudes and the absolute magnitudes of the host galaxy and the polar structure.

\begin{deluxetable*}{lccccc}
  \tablenum{3}
  \tablecaption{Estimated magnitudes.\label{tab:magnitude}}
  \tablewidth{0pt}
  \tablehead{
   \colhead{} & \colhead{$g$} & \colhead{$r$} & \colhead{$i$} & \colhead{$z$} & \colhead{$y$} 
  }
  \startdata
     $m_{\rm HG}$ & $20.07$ & $19.14$ & $18.67$ & $18.29$ & $18.17$ \\ 
     $m_{\rm PR}$ & $21.02$ & $20.41$ & $20.00$ & $20.01$ & $19.76$ \\
     $M_{\rm HG}$ & $-20.30$ & $-20.92$ & $-21.34$ & $-21.73$ & $-21.82$ \\
     $M_{\rm PR}$ & $-19.09$ & $-19.52$ & $-19.85$ & $-19.88$ & $-20.22$ \\
  \enddata
  \tablecomments{
    \hangindent6pt\noindent
    Note that $m$ and $M$ are the apparent and absolute magnitudes, respectively. 
    \hangindent6pt\noindent
    HG and PR are  the host galaxy and the polar structure, respectively. 
    \hangindent6pt\noindent
    We note that the photometric errors in all the five bands are less than 0.01 mag.
    \hangindent6pt\noindent
    The absolute magnitudes are corrected for the Galactic extinction using the values listed in the HSC-SSP PDR2 data,
    which are estimated from the dust maps given in  \citet{schlegel_maps_1998} ; see also \cite{aihara_second_2019}. 
    \hangindent6pt\noindent
    The $k$-correction is applied for all the absolute magnitudes  (\cite{chilingarian_analytical_2010}).    
  }
  \end{deluxetable*}

The spectral energy distribution (SED) of the galaxy is a composite of the spectral emissions from all the stars contained therein. 
Therefore, we can estimate  the stellar mass of  both the host galaxy and the polar structure.

Here, we use the absolute magnitudes in $i$-band and the color $g - i$ following the method outlined by \citet{taylor_galaxy_2011}, and obtain the stellar mass of the host galaxy and the polar structure are $26.18 \times 10^9 M_\odot$ and $4.23 \times 10^9 M_\odot$, respectively.

The characteristic absolute magnitude $M_*$ corresponding to the characteristic luminosity $L_*$ of the galaxies are suggested from \citet{blanton_luminosity_2001} as follows; $u^{\prime}=-18.34\pm 0.08$ mag, $g^{\prime}=-20.04\pm 0.04$ mag, $r^{\prime}=-20.83\pm 0.03$ mag, $i^{\prime}=-21.26\pm 0.04$ mag, and $z^{\prime}=-21.55\pm 0.04$ mag;
see table 1 of \citet{blanton_luminosity_2001} for details on the sample galaxies.

Using the color conversion formulae from the SDSS system to the HSC system from \citet{komiyama_stellar_2018}, these $M_*$ are converted to $g=-20.12$ mag and $i=-21.31$ mag.
The $g-$ and $i-$ band absolute magnitudes of the J0953 host galaxy are -20.30 mag and -21.34 mag, respectively.
Therefore, the host galaxy of J0953 is possibly brighter and heavier than the galaxies of the characteristic luminosity $L_*$ corresponding to the characteristic absolute magnitude $M_*$. 
 
\begin{figure}
   \begin{center}
      \includegraphics[width=80mm]{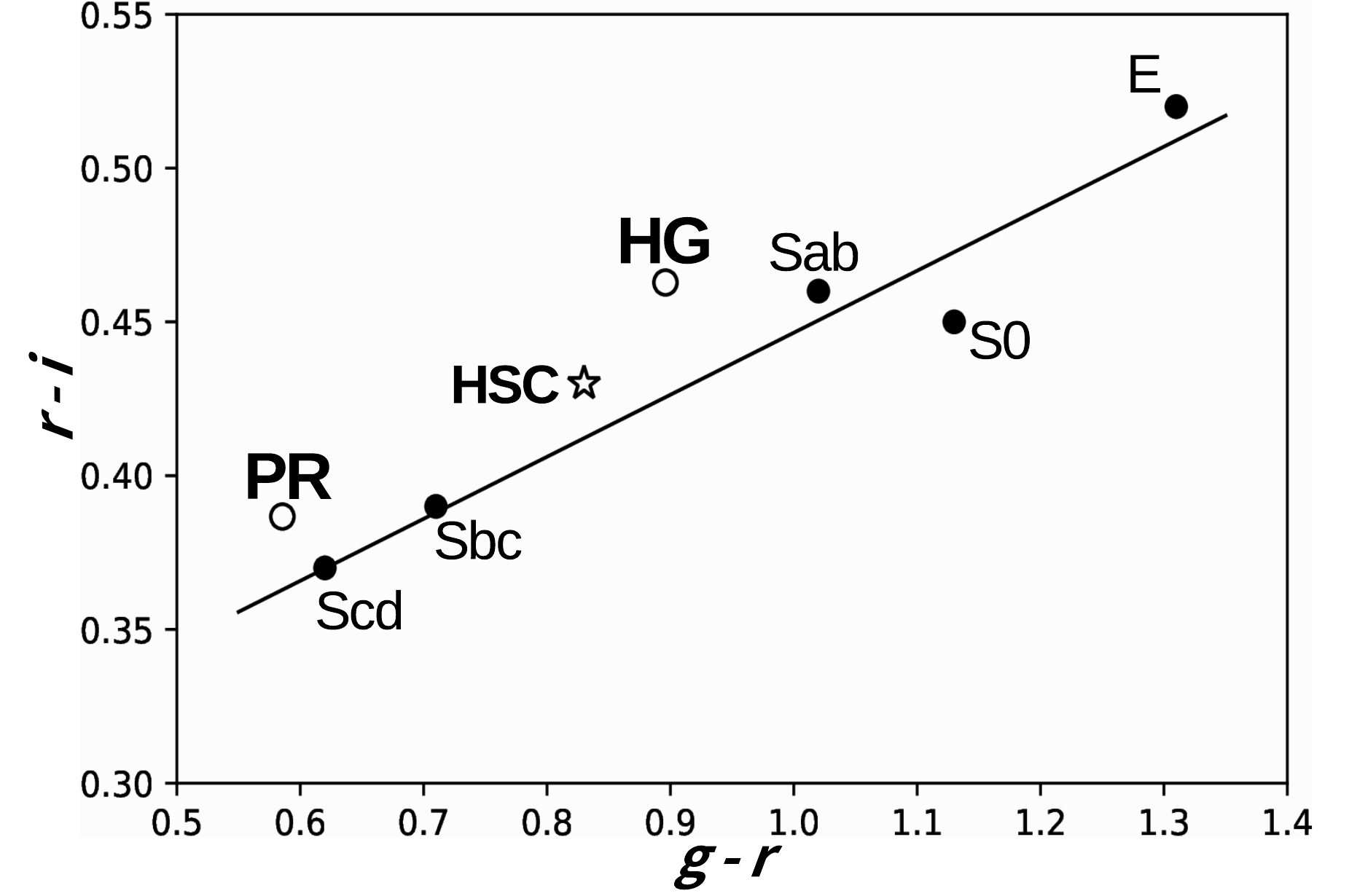}
   \end{center}
   \caption{The $r -i$ vs, $g - r$  diagram. 
   The filled circles show the colors  for 5 morphology types (E, S0, Sab, Sbc, and Scd) of galaxies in \citet{fukugita_galaxy_1995}. 
   The solid line shows the best linear fit to the colors of 5 morphology types. 
   The open circles of HG and PR show the colors of the host galaxy and the polar structure of J0953. 
   The star mark with HSC shows the colors taken  from table \ref{tab:data} of J0953.
   All of the above colors are at $z \sim 0.2$ and corrected for the Galactic extinction.}
   \label{fig:color}
\end{figure}

Figure \ref{fig:color} is the color-color diagrams to estimate one of the information, and the colors of the host galaxy and the polar structure of this galaxy at $z \sim 0.20$ are plotted. 

Figure \ref{fig:color} is the color-color ($r -i$ vs.~$g - r$) diagram 
of the host galaxy and the polar structure compared with
that of each Hubble type of typical galaxies at $z \sim 0.2$ (\cite{fukugita_galaxy_1995}).
The colors of the host galaxy suggest a Sab-Sb galaxy while host galaxies of typical PRGs are 
mostly early-type galaxies (E/S0) (\cite{reshetnikov_new_2019}).
The polar structure is similar in colors to Scd galaxies.
This is consistent with that the polar structure is blue and probably younger than the host galaxy.
In order to confirm if this galaxy is truly a PRG, it is necessary to make spectroscopic observations,
to investigate the kinematical properties of both the host galaxy and the polar structure.

\section*{Acknowledgements}

  We would like to thank an anonymous referee for his/her useful comments.
  The HSC collaboration includes the astronomical communities of Japan and Taiwan, and Princeton University. The HSC instrumentation and software were developed by the National Astronomical Observatory of Japan (NAOJ), the Kavli Institute for the Physics and Mathematics of the Universe (Kavli IPMU), the University of Tokyo, the High Energy Accelerator Research Organization (KEK), the Academia Sinica Institute for Astronomy and Astrophysics in Taiwan (ASIAA), and Princeton University. Funding was contributed by the FIRST program from Japanese Cabinet Office, the Ministry of Education, Culture, Sports, Science and Technology (MEXT), the Japan Society for the Promotion of Science (JSPS), Japan Science and Technology Agency (JST), the Toray Science Foundation, NAOJ, Kavli IPMU, KEK, ASIAA, and Princeton University. 
  This paper makes use of software developed for the Large Synoptic Survey Telescope. We thank the LSST Project for making their code available as free software at  $\langle $http://dm.lsst.org$\rangle $. 
  The Pan-STARRS1 Surveys (PS1) have been made possible through contributions of the Institute for Astronomy, the University of Hawaii, the Pan-STARRS Project Office, the Max-Planck Society and its participating institutes, the Max Planck Institute for Astronomy, Heidelberg and the Max Planck Institute for Extraterrestrial Physics, Garching, The Johns Hopkins University, Durham University, the University of Edinburgh, Queen's University Belfast, the Harvard-Smithsonian Center for Astrophysics, the Las Cumbres Observatory Global Telescope Network Incorporated, the National Central University of Taiwan, the Space Telescope Science Institute, the National Aeronautics and Space Administration under Grant No. NNX08AR22G issued through the Planetary Science Division of the NASA Science Mission Directorate, the National Science Foundation under Grant No. AST-1238877, the University of Maryland, and Eotvos Lorand University (ELTE) and the Los Alamos National Laboratory.
  This paper is based on data collected at the Subaru Telescope and retrieved from the HSC data archive system, which is operated by Subaru Telescope and Astronomy Data Center (ADC) at NAOJ. Data analysis was in part carried out on the Multi-wavelength Data Analysis System operated by ADC.


\bibliography{PASP_220824}{}
\bibliographystyle{aasjournal}



\end{document}